# Plug-Play Plasmonic Metafibers for Ultrafast Fiber Lasers

*Lei Zhang[1,2,3#], Huiru Zhang[4#], Ni Tang[1,2,3], Xiren Chen[4], Fengjiang Liu[2,3], Xiaoyu Sun[1,2,3], Hongyan Yu[1,2,3], Xinyu Sun[1,2,3], Qiannan Jia[1,2,3], Boqu Chen[1,2,3], Benoit Cluzel[5], Philippe Grelu[5], Aurelien Coillet[5], Feng Qiu[2,3], Lei Ying[6], Wei E. I. Sha[7], Xiaofeng Liu[8], Jianrong Qiu[1], Ding Zhao[2,3], Wei Yan[2,3], Duanduan Wu[4\*], Xiang Shen[4\*], Jiyong Wang[2,3\*], Min Qiu[2,3\*]*

[1] State Key Laboratory of Modern Optical Instrumentation, College of Optical Science and Engineering, Zhejiang University, Hangzhou 310027, P.R. China

[2] Key Laboratory of 3D Micro/Nano Fabrication and Characterization of Zhejiang Province, School of Engineering, Westlake University, 18 Shilongshan Road, Hangzhou 310024, Zhejiang Province, P.R. China

[3] Institute of Advanced Technology, Westlake Institute for Advanced Study, 18 Shilongshan Road, Hangzhou 310024, Zhejiang Province, P.R. China

[4] Laboratory of Infrared Material and Devices & Key Laboratory of Photoelectric Materials and Devices of Zhejiang Province, Advanced Technology Research Institute, Ningbo University, Ningbo 315211, P.R. China

[5] Laboratoire Interdisciplinaire Carnot de Bourgogne, UMR 6303 Université Bourgogne Franche-Comté, 9 avenue Alain Savary, 21078 Dijon, France

[6] Interdisciplinary Center for Quantum Information and Department of Physics, Zhejiang University, Hangzhou, P. R. China

[7] College of Information Science and Electronic Engineering, Zhejiang University, Hangzhou 310027, P. R. China

[8] School of Materials Science and Engineering, Zhejiang University, Hangzhou 310027, P.R. China

# Equal contributions;

*Corresponding authors: qiumin@westlake.edu.cn; wangjiyong@westlake.edu.cn; wuduanduan@nbu.edu.cn; shenxiang@nbu.edu.cn

## Abstract

Metafibers expand the functionalities of conventional optical fibers to unprecedented nanoscale light manipulations by integrating metasurfaces on the fiber tips, becoming an emerging light-coupling platform for both nanoscience and fiber optics communities. Mostly exploring the isolated bare fibers, current metafibers remain as proof-of-concept demonstrations due to a lack of standard interfaces with the universal fiber networks. Here, we develop new methodologies to fabricate well-defined plasmonic metasurfaces directly on the end facets of commercial single mode fiber jumpers using standard planar technologies and provide a first demonstration of their practical applications in the nonlinear optics regime. Featuring plug-play connections with fiber circuitry and arbitrary metasurfaces landscapes, the metafibers with tunable plasmonic resonances are implemented into fiber laser cavities, yielding all-fiber sub-picosecond (minimum 513 fs) soliton mode locked lasers at optical wavelengths of 1.5 μm and 2 μm, demonstrating their unusual polarimetric nonlinear transfer functions and superior saturation absorption responses. Novel insights into the physical mechanisms behind the saturable absorption of plasmonic metasurfaces are provided. The nanofabrication process flow is compatible with existing cleanroom technologies, offering metafibers an avenue to be a regular member of functionalized fiber components. The work paves the way towards next generation of ultrafast fiber lasers, optical frequency combs, optical neural networks and ultracompact "all-in-fibers" optical systems for sensing, imaging, communications, and many others.

**Introduction**

Plasmonic metasurfaces are formed by artificially patterning the subwavelength nanostructures in a two dimensional (2D) array.[1-4] Their plasmonic resonances are intendedly designed by altering the spatial distribution and orientation of unit cells, in order to improve the interactions of the impinging light through enhanced the absorption and scattering cross-sections, both in linear and nonlinear optical regimes.[5-10] Indeed, the nonlinear absorption of plasmonic metasurfaces is arousing more and more interest in recent years for the state-of-art applications ranging from frequency conversion,[9, 10] neuromorphic circuits [11-13] to ultrashort laser pulse generation.[14, 15] Pioneer works using plasmonic metasurfaces as saturable absorbers(SAs) in a fiber laser operating in soliton mode-locking regimes.[15, 16] Owning the superiorities of tunable modulation depths and a strong bonding to the plasmonic nature over colloidal nanoparticles with dispersed sizes and random orientations extensively investigated in fiber lasers,[14, 17, 18] the metasurfaces SA was, however, solely demonstrated in a free-space coupled fiber laser cavity, regardless of practical packaging/integrating requirements and physical mechanisms of saturable absorption.

Plasmonic metasurfaces are routinely fabricated on the planar wafers, maintaining a large compatibility with conventional semiconductor technologies. Advanced nonplanar substrates involving atomic force microscope tips, carbon nanotubes and silver nanowires have been explored in recent years with state-of-art nanofabrication methods, [19, 20] which are limited to sole scientific demonstrations. The optical fiber tips combining the fiber optics and planar technologies is, however, emerging as one of the most promising light-coupling platforms for both scientific and industrial communities.[21, 22] Integrating the plasmonic metasurfaces on the optical fiber tips forming the so-called metafibers expands the functionalities of an ordinary optical fiber to a nanoscale manipulation of light, yielding a variety of advanced applications such as ultracompact sensing,[23] super-resolution imaging,[24] and planar waveshaping.[25] There are several strategies to fabricate the metafibers by using either standard planar technologies, e.g. focused ion beam (FIB),[22] electron-beam lithography (EBL),[26, 27] and nanoimprint,[28, 29] or unconventional in-situ approaches, i.e. ice lithography.[19, 23] However, up to now, plasmonic metafibers have predominantly explored separate bare fibers without practical applications in nonlinear optics regimes. There are also certain challenges for a widespread uptake of metafibers as regular component devices for fiber optics: a) The nanofabrication suffers from inevitable mechanical vibrations and thus a poor repeatability of nanostructures due to a large aspect ratio of bare fibers; b) The connections between the functionalized metafibers and standard optical fibers introduce either potential contaminations to the plasmonic metasurfaces or considerably high insertion losses within the fiber circuit. Thus, methods to fabricate the metafibers with a reproducible geometry of metasurfaces as well as standard adapting interfaces are clearly needed.

In this work, we propose new methodologies that integrate the well-defined metasurfaces directly on the endfaces of commercial single mode fiber jumpers (SMFJs), as shown in Fig. 1(a), by using the standard planar technologies. Widely spread in the fiber-optics community and industry, SMFJs hold the fibers stably in their central positions and provide connection interfaces with the ordinary fiber networks, e. g. physical contact, ultra-physical contact and angled physical contact, better depressing the back reflection of the light and thus reducing the return loss. To demonstrate the low insertion loss and a practical application in nonlinear plasmonics, the metafibers using SMFJs are implemented into fiber laser cavities, serving as special SAs that filter out the quasi-continuous wave (cw) laser background while promoting ultrashort pulse emission within the laser cavity (Fig. 1(a)). By tuning the plasmonic resonances of the metafibers, we are able to achieve all-fiber sub-picosecond soliton mode-locking working at different optical wavelength bands. In the end, insights into the physical mechanisms behind the saturable absorption of plasmonic metasurfaces are given.

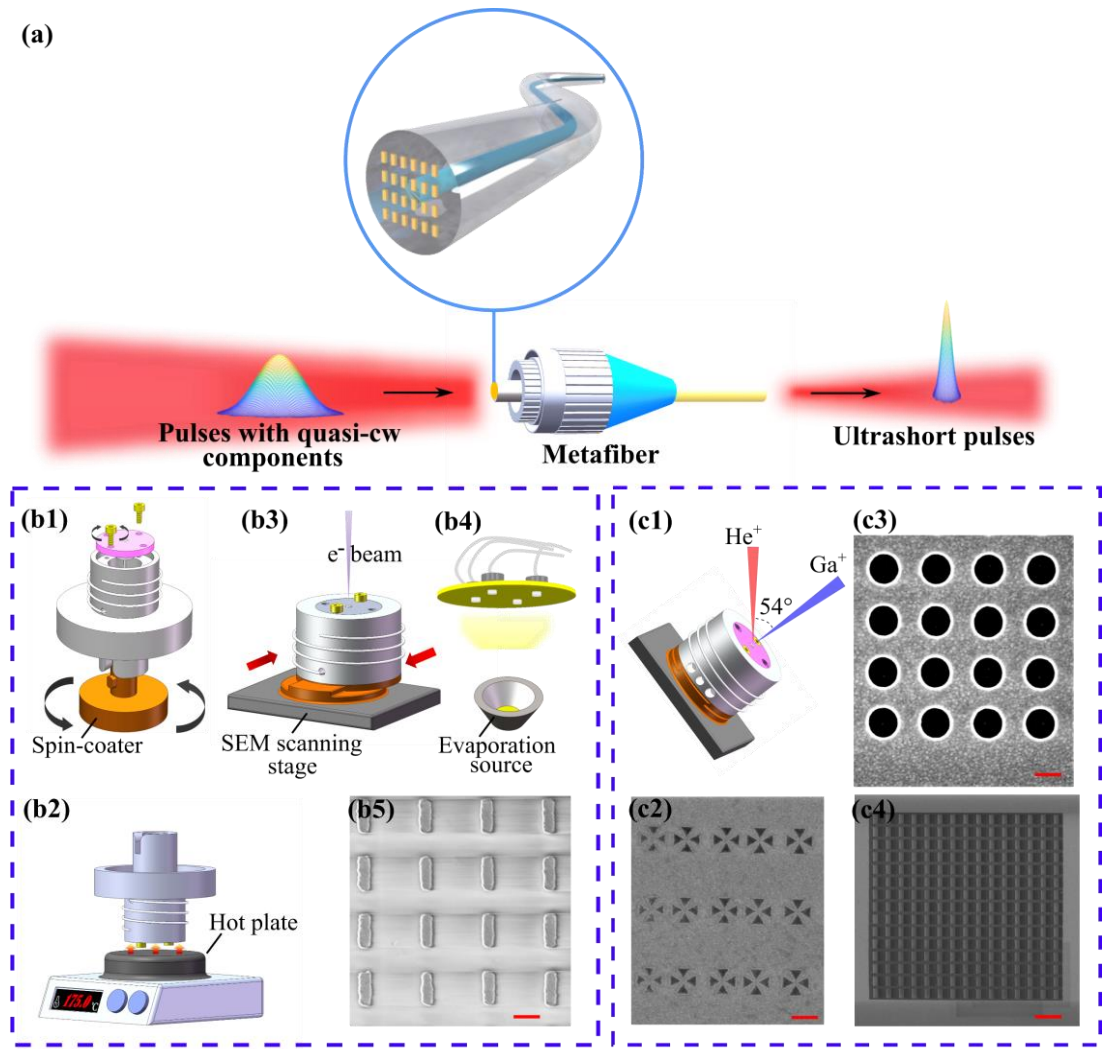

Figure 1. **Nanofabrication of the plasmonic metafibers.** (a) Schematic diagram of plasmonic metasurfaces integrated on the endface of a standard SMFJ used here as SAs for laser mode locking. (b1-b5) Nanofabrication of plasmonic metafibers by using standard EBL. (b1) Home-built rotating chunk for the spin-coating of electron-beam resists. (b2) Alternative use of the rotating chunk for the soft-baking. (b3) Translating chunk for the electron-beam exposure. The bottom part is adapted to the scanning stage of a commercial SEM. (b4) SMFJ holder for the physical vapor deposition. (b5) The SEM image of an Au nanorod array fabricated on the endface of a SMFJ. The scale bar is 400 nm. The height of the nanostructures is 50 nm. The color contrast in the substrate is caused by the charge accumulations during the SEM observation. (c1-c4) Nanofabrication of plasmonic metafibers by using standard FIB. (c1) Translating chunk for the ion-beam exposure, where $He^+$ beam is used for imaging and the $Ga^+$ beam is for milling. The translating chunk is tilted at an angle of 54° in horizontal direction. (c2-c4) The SEM images of orthogonal Au bow-ties, Au nanoholes and Au nanorod arrays fabricated on the endface of a SMFJ. The scale bars in (c2-c4) represent 1 μm, 400 nm and 2 μm, respectively. The height of the nanostructures is 60 nm. The horizontal lines connecting two rows of nanorods in c4 are caused by the charge accumulations during the SEM observations.

## Results

In order to pattern the plasmonic metasurfaces on the endfaces of commercial SMFJs, the first planar technology we employ is the standard EBL, associated with various customized mechanical parts, as it is shown in Figs. 1(b1~b5). A SMFJ (working at 980-1650 nm, FC/PC, Ø900 µm jacket, 1 m long) is cut into two from the middle and one of which is stably mounted on a fiber adapter. As illustrated in Fig. 1(b1), the fiber adapter can be connected to a rotating chunk designed to fit a commercial spin-coater (SUSS MicroTec). After the spin-coating of the electron sensitive resist, the rotating chunk is flipped top down and placed on a hot plate for soft-baking, as shown in Fig. 1(b2). The spacing (2.5 mm) between the SMFJ endface and the top surface of the hot plate is guaranteed by the heads of the four screws used for mounting the fiber adapter onto the chunk. Then, the fiber adapter is placed in a commercial scanning electron microscope (SEM, Zeiss Crossbeam 550) for the electron-beam exposure, thanks to a home-built translating chunk mounted on the scanning stage of the SEM, as shown in Fig. 1(b3). After the exposure and development, the patterned SMFJ is transferred to an evaporator via a home-built evaporation adaptor for depositing the target materials, as illustrated in Fig. 1(b4). A final lift-off reveals the metasurface on the fiber endface, concluding the metafiber fabrication process flow. Fig. 1(b5) shows a typical metasurface made of a nanorod array fabricated on a SMFJ endface by using this technique (see "Nanofabrication of the metafiber by using EBL" from SI for details). The advantages of such method include low cost and high immunity to complicate plasmonic systems, such as the heterodimers for instance. [10, 30, 31]

Alternatively, we also demonstrate the possibility to fabricate plasmonic meta-surfaces on the SMFJ endface using the standard FIB milling, as illustrated in Figs. 1(c1~c4). The same evaporation adaptor shown in Fig. 1(b4) is firstly used to deposit the target metal, e. g. Au. The SMFJ is then transferred to a home-built translating chunk mounted on the scanning stage of a commercial FIB instrument (Carl Zeiss, ORION NanoFab, USA) for the milling, as shown in Fig. 1(c1). Fig. 1(c2~c4) shows the SEM images of three kinds of plasmonic metasurfaces on the endfaces of standard SMFJs after the milling: orthogonal bow-ties, nanoholes and nanorods, respectively (see "Nanofabrication of the metafiber by using FIB" from SI for details). In comparison with EBL, FIB needs much fewer preparing procedures and provides a relatively better resolution for nanopatterning. In addition, since resist spin-coating is not needed, the FIB process flow is compatible with nonplanar fiber interfaces, e. g. angled physical contact, which offers a fast and convenient way to functionalize most of the optical fibers available commercially.

Thanks to our versatile nanomanufacturing setups, the metasurfaces with specific plasmonic properties can be easily transferred from planar substrates to the fiber endfaces. We have thus employed these techniques to fabricate saturable metafibers and provide a demonstration of their use for laser mode-locking. First, to achieve a saturation of optical absorption from a metasurface, we need to tune the plasmonic resonance to the target wavelength, e. g. the C+L telecommunication optical wavebands at 1.5 µm which can be entirely covered simultaneously thanks to the broad resonance of the same plasmonic metasurface. [15] Nanorods are one kind of the most common nanostructures to achieve the designed plasmonic functionalities. [15, 32] Indeed, their resonances could be easily tuned from the visible to the near-infrared regimes by altering their aspect ratios. [15, 32] Fig. 2(a) shows the theoretical transmission spectra of a 50 nm thin metasurfaces made of Au nanorods arranged in a square array with a 750 nm period, a constant rod width of 160 nm and varying lengths. (see "Linear optical response calculations" from SI for details). The excitation polarization is parallel to the long axis of nanorods. From Fig. 2(a), the metasurface resonates at 1550 nm, the central wavelength of telecommunication band, for a rod length of 470 nm. Such a resonance directly comes from the longitudinal dipolar plasmonic mode of the individual nanorods. According to these numerical calculations, we fabricate the corresponding plasmonic metasurfaces on the endface of a standard SMFJ by using EBL, as shown in Fig. 2(b). The measured lengths and the widths of the nanorods are 485 nm and 155 nm respectively, with a tolerance of 10 nm. The inset shows the fundamental electric field distribution of a nanorod array with these actual dimensions excited by a plane wave of 1550 nm. The incident polarization is linear and parallel to the long axis of the nanorods. Strongly confined hot spots can be clearly observed in the near field of the nanorods, see the inset of Fig. 2(b), emphasizing on the excitation of the longitudinal dipolar modes.

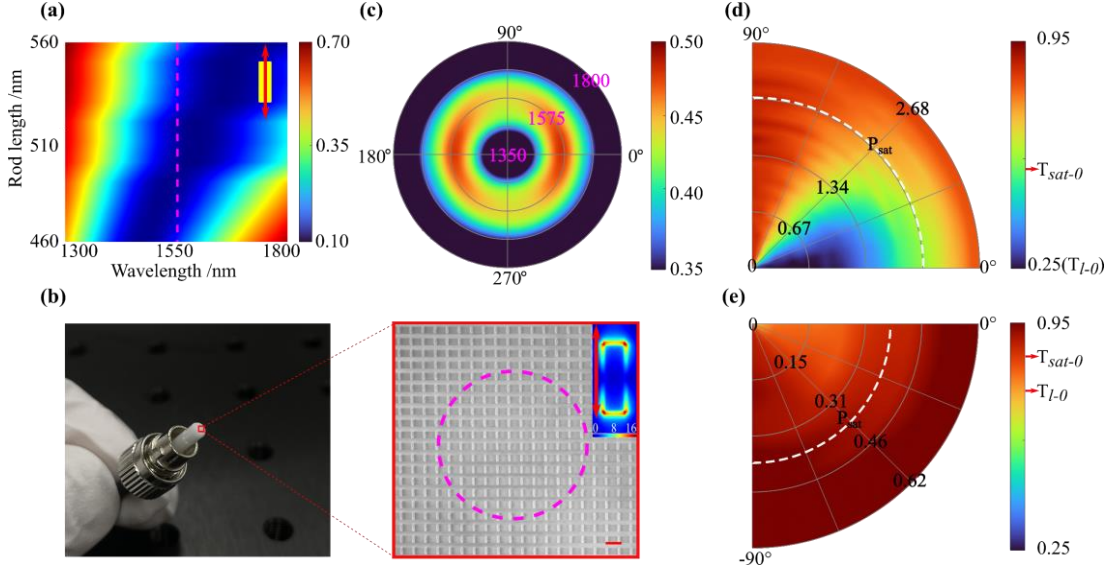

**Figure 2. Nanocharacterization of the plasmonic metafibers.** (a) Calculated transmission spectra of the Au nanorod arrays with varying rod lengths. The rod width is 160 nm and the height is 50 nm. The longitudinal and transverse periods remain constant at 750 nm. The excitation polarization is parallel to the long axis of the nanorods. The pink dashed line indicates the spectral wavelength of 1550 nm. (b) A photograph of the fabricated metafiber by using EBL and the corresponding SEM image in the fiber core region. The scale bar represents 800 nm. The dashed pink circle indicates the boundary of the fiber core. The inset shows the fundamental electrical field distribution of the nanorod array (taking periodic conditions into account). The red arrow represents the incident polarization. (c) Experimental polarimetric extinction spectra of the plasmonic metafiber. The polar coordinates ($\lambda$, $\theta$) represent the spectral wavelength and the polarization angle of the incident light. The color contrast represents the extinction level of the metafiber. (d~e) Power- and polarization-dependent nonlinear transmission of a nanorod array in on-resonance (using 1550 nm fs laser) and off-resonance (using 1950 nm fs laser) excitation conditions. The polar coordinates ($P$, $\theta$) represent the averaged power in the focus and the polarization angle of the incident light. The color contrast represents the transmission level of the nanorod array. The white dashed curves represent averaged saturation powers. $T_{l\text{-}0}$ and $T_{sat\text{-}0}$ denote the transmission levels at $\theta=0°$ in the linear absorption regime and once absorption is saturated, respectively.

In order to characterize the optical responses, i. e. the resonant absorptions, of the metafibers, a home-built extinction microscope is firstly employed (see "Optical setups for measuring the extinction spectra" from SI for details). A supercontinuum light source (SLS, NKT photonics) emitting in the 1~1.8 μm spectral range is used to excite the plasmonic metasurfaces. The measured extinction spectra with the polarization dependence are shown in Fig. 2(c). The polar coordinates ($\lambda$, $\theta$) represent the spectral wavelength and the incident polarization angle. $\theta = 0°$ corresponds to an incident polarization parallel to the long axis of the nanorods. The color contrast represents the extinction level of the metafiber. The extinction spectrum, which denotes the fingerprint of plasmonic resonance, is defined as Ext ($\lambda$)= [ $T_{ref}$ ($\lambda$) -$T_{nr}$($\lambda$)]/$T_{ref}$ ($\lambda$), where $T_{ref}$ ($\lambda$) is the transmission of a reference fiber, $T_{nr}$ ($\lambda$) is the transmission of the metafiber, and $\lambda$ is the wavelength. The reference fiber is identical with the metafiber except having no plasmonic nanostructures on the endface. The dipole-radiation-like map demonstrates the dominant roles of longitudinal dipolar mode of the metasurfaces. The plasmonic resonances generally locate at 1550 nm and have an averaged full-width at half maximum (FWHM) of 200 nm. Depending on the excitation polarization, the extinction fluctuates in an approximate sine/cosine function.

Following the linear optical characterization, the nanorod metasurfaces with the same patterns are fabricated on a glass substrate with the same refraction index as the SMF core by using standard EBL, in order to further investigate their power- and polarization-dependent nonlinear optical responses. The nonlinear transmittance is measured using a home-built optical setup with an ultrashort pulsed pumping laser (see "Optical setups for measuring nonlinear optical transmissions" from SI for details). The repetition rate of the laser is 10.32 KHz and the output wavelength can be tuned from 600 nm to 2400 nm via an optical parametric amplifier. The pulse duration is slightly varied with the output wavelength, e. g. changing from 179 fs to 166 fs as the wavelength is tuned from 1550 nm to 1950 nm. The incident polarization is linear, such that the polarization axis with respect to the nanorod orientation is tuned with a half-waveplate. Using this laser, we record the transmission of the nanorod array under study and use the transmission of the nearby blank glass slide as a reference.

We firstly use the excitation laser at the central wavelength of 1550 nm, where the plasmonic resonance of the nanorod array locates. The result is plotted in a polar pseudocolor diagram shown in Fig. 2(d), in which the polar coordinates ($P$, $\theta$) represent the averaged power in the focus and the input polarization angle, and the colormap represents the transmission level of the array. $\theta$ ranges from 0° to 90° with respect to the long axis of the nanorods. With this representation, the saturable absorption with the power- and polarization-dependencies clearly appears. At the lower power, the transmission of the nanorod array remains a constant, $T_l$, for a given polarization. In this linear regime, the transmission strongly depends on the incident polarization since $T_l$ is inversely proportional to the absorption cross-section of the nanorod. Above a critical power, the overall transmission increases nonlinearly until it reaches a saturated value $T_{sat}$, showing the saturation of the metasurface absorption. For all input polarization orientation, the power-dependent transmission coefficient shows a 'S'-shape profile, from which the modulation depth, saturation intensity $P_{sat}$ and the parameters of $T_l$ and $T_{sat}$ could be obtained by fitting the data. (see "Laser intensity dependent optical transmission model" from SI for details) The modulation depth shows a strong dependence to the incident polarization, ranging from 5% to 57% when the polarization is tuned from the short axis to the long axis of nanorods. Such high modulation depths overweight the best performance of 2D-saturable absorbers. [33] When the same nanorod metasurface is excited with a 1950 nm pulsed laser, a clear decrease of polarization dependency is observed from Fig. 2(e). This lowering of the polarization dependency also comes with a global growth of the transmission at the minimum 70% and a drop of the modulation depth at the maximum 26%. The mismatch between the plasmonic resonance and the fundamental excitation will significantly reduce the linear and nonlinear absorptions of nanorod array, resulting in an unremarkable saturable absorption. The results in Figs. 2(d~e) can probably explain the reason why considerable low modulation depths were reported previously on colloidal Au nanorods, as the dispersed sizes and orientations were averaging the saturable absorption and the critical contribution of plasmonic resonances. [14, 17, 18]

The saturable absorption can be helpful for achieving the formation of ultrashort pulses in laser architectures, ultimately reaching self-starting passively mode locked regimes. [15] To further test the fabricated samples and provide a practical demonstration of the saturable metasurfaces application in nonlinear optics, the metafibers are integrated into the fiber laser cavities to promote mode locking. The fiber cavity shown in Fig. 3 is built, which includes a 980 nm pump diode (LD, maximum pump power of 88 mW), a 980/1550 nm wavelength multiplexer (WDM), 35 cm of erbium-doped fiber (EDF, Nufern, SM-ESF-7/125), a polarization-insensitive isolator (ISO), a polarization controller (PC), and an output fiber coupler (OFC). The PC is used to alter the laser polarization in the cavity, targeting a proper SA efficiency in virtue of the polarimetric properties of metafibers shown in Figs. 2(c~d). The coupler extracts 10% of the laser energy for the pulse characterization. The overall length of the cavity is 8.8 m with anomalous net chromatic dispersion and all fiber connections are made with standard telecom fiber SMF-28e.

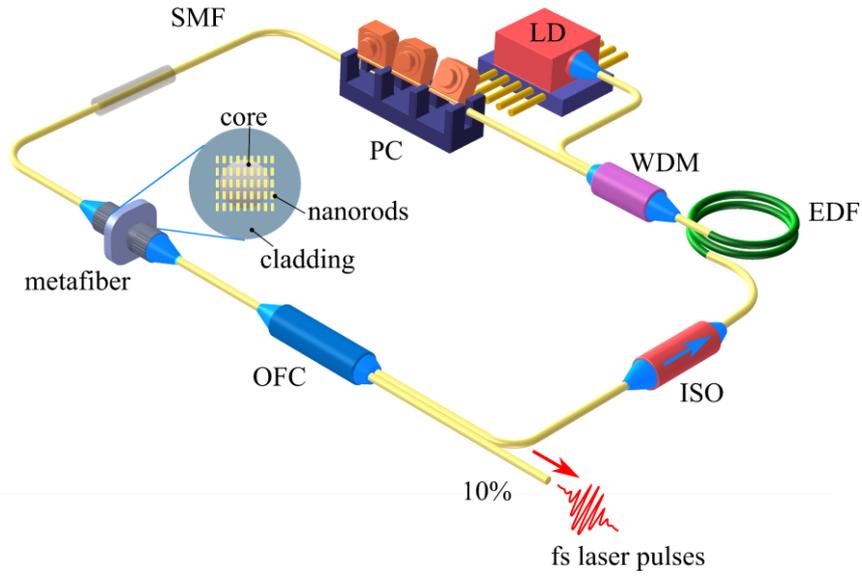

**Figure 3**. **Home-built ultrafast fiber laser integrating a plasmonic metafiber**, where LD represents the pump laser diode, WDM is the wavelength-division multiplexer, EDF is the erbium-doped fiber, ISO is the optical isolator, OFC is the output fiber coupler, SMF stands for standard single-mode telecom fiber, PC is the polarization controller.

In such an overall anomalous dispersion regime, the chromatic dispersion and the self-phase modulation accumulated during a cavity roundtrip can balance on average, preventing a significant pulse broadening and leading to a so-called optical soliton regime. [34] The output pulse train and radio frequency (RF) spectra are monitored by an oscilloscope (Tektronix, 1 GS/s, 100 MHz) and a RF spectrum analyzer (Keysight, N9000B) respectively. The pulse spectra, output power and autocorrelation trace are monitored by an OSA (Yokogawa, AQ6375), a power meter (Thorlabs, S148C) and an autocorrelator (FR-103XL) respectively. Fig. 4(a) shows the average laser output power featuring a cw laser conversion efficiency of 5.5% without the metafiber SA, noting that the conversion efficiency is limited by the moderate 10% output coupling. The laser threshold is ~13 mW, proving the relatively low loss of the whole cavity. After implementation of the metafiber SA into the laser cavity, the laser threshold increases to ~23 mW due to the insertion loss, and the conversion efficiency of the laser reduces to 1.8%. We note the low mode locking threshold at a pump power of 38 mW, resulting from an efficient laser regime discrimination performed by the metafiber SA. Fig. 4(b) shows the optical spectrum of the laser mode-locked with the metafiber SA, for a pump power of 49 mW. The spectrum has a smooth broadband central part with a 3-dB bandwidth of 6 nm, centered at 1560 nm. Gordon-Kelly sidebands appear on both sides of the spectrum, which is a typical feature of fiber lasers mode-locked in the anomalous dispersion regime. [35] Fig. 4(c) shows the optical autocorrelation trace at the pump power of 58 mW. A pulse duration of 513 fs is inferred from the measurement, and the time-bandwidth product (TBP) is estimated to be 0.379, indicating that the output pulses are just slightly chirped, as compared with the expected TBP of 0.315 for an unchirped exact hyperbolic-secant shaped pulse. [34] The time interval between subsequent output pulses is measured to be 44 ns, which corresponds to the fundamental frequency of 22.7 MHz in Fig. 4(d). Fig. 4(d) shows the RF spectra of the soliton mode-locked pulse at the pump power of 58 mW. The signal-to-noise ratio of the fundamental frequency of the laser can reach 75 dB at a resolution of 300 Hz, and the signal-to-noise ratio of the RF spectrum in the range of 0-1 GHz is greater than 60 dB at a resolution of 10 kHz, indicating a remarkably high stability of the laser pulses.

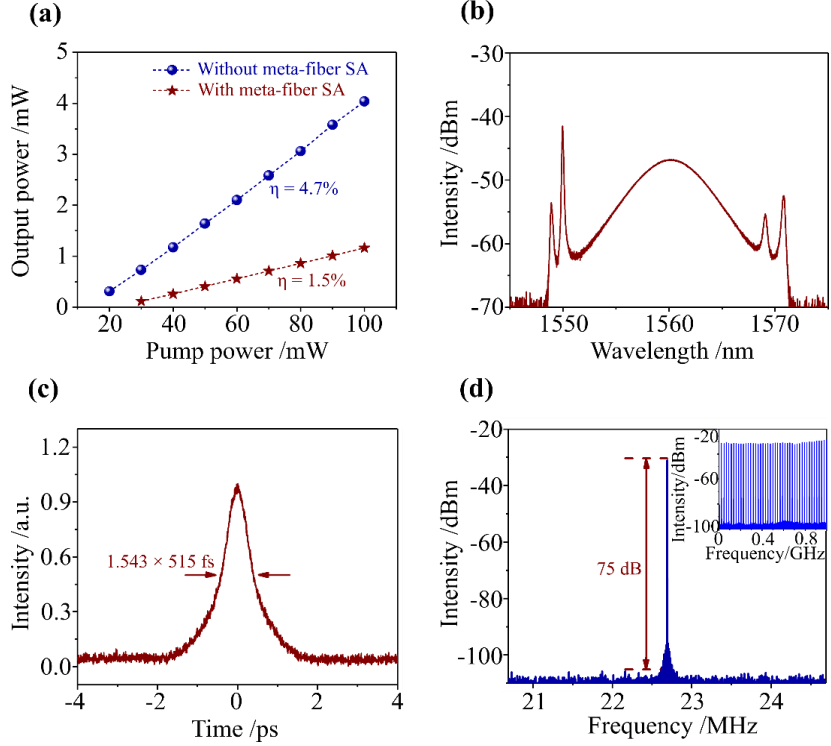

**Figure 4. Soliton mode-locking at 1.5 μm optical waveband incorporated with a plasmonic metafiber.** (a) Averaged output power of a home-buit 1.5 μm fiber laser with (red stars) and without (blue dots) incorporating a metafiber SA as functions of the pump power. (b) The soliton pulse spectrum at the pump power of 49 mW. (c) Autocorrelation trace of a single soliton at the pump power of 58 mW. (d) RF spectrum of the soliton pulses with a scanning range of 5 MHz and a resolution of 300 Hz. Inset shows the RF spectrum of the soliton pulses with a scanning range of 1 GHz and a resolution of 10 KHz.

The soliton pulses can be tuned to other optical wavelength bands, e. g. 2 μm in a thulium fiber laser, by adapting the plasmonic resonance of a metafiber to the gain spectrum of the laser cavity. To achieve a resonant absorption at 2 μm for example, we use the similar nanorod metasurfaces but with different landscapes. As shown in Fig. 5(a), the transmission spectra of the nanorod arrays with constant width (340 nm) and spacings (longitudinal and transverse periods: 1μm) but different lengths are calculated. The rod length can be evaluated as ~680 nm, in which the longitudinal dipolar mode of the plasmonic metasurfaces is located at 2 μm. Guided by the numerical predictions, we fabricate the corresponding plasmonic metasurfaces on a SMFJ endface by using FIB, as shown in the SEM image of Fig. 5(b). The longitudinal and transverse periods keep the same: 1μm. The length and the width of nanorods are measured as 690 nm and 345 nm with a tolerance of 10 nm, respectively. The inset shows the fundamental electric field distribution of a nanorod array with the actual dimensions excited by a linear polarized plane wave at 2 μm. The nonlinear transmission spectra with incident power- and polarization-dependences are measured in both on-resonance and off-resonance cases, similarly as Figs. 2(c, d). In on-resonance case, fs-pulsed laser with the central wavelength of 1950 nm is used as the light excitation source (Fig. 5(c)), while in off-resonance case the wavelength of the excitation laser is switched to 1550 nm (Fig. 5(d)). Similarly, a strong polarimetric linear and nonlinear absorption could be obviously observed from the former, while the polarization dependency almost disappears for the latter. The maximum modulation depth in the former case is 37%, while it decreases to only 18% for the latter. The metafiber is then placed into a laser cavity operating in 2 μm region, which results in a stable soliton mode-locking (see "Soliton mode-locking at 2 μm" from SI for details). The time-averaged pulse spectrum is shown in Fig. 5(e) at the pump power of 310 mW, featuring multiple pairs of symmetric sidebands. The central wavelength is located at 1921.1 nm, and the FWHM of the spectrum is 3.2 nm. Fig. 5(f) shows the pulse autocorrelation trace and pulse trains at the pump power of 310 mW. A pulse duration of 1.42 ps is measured.

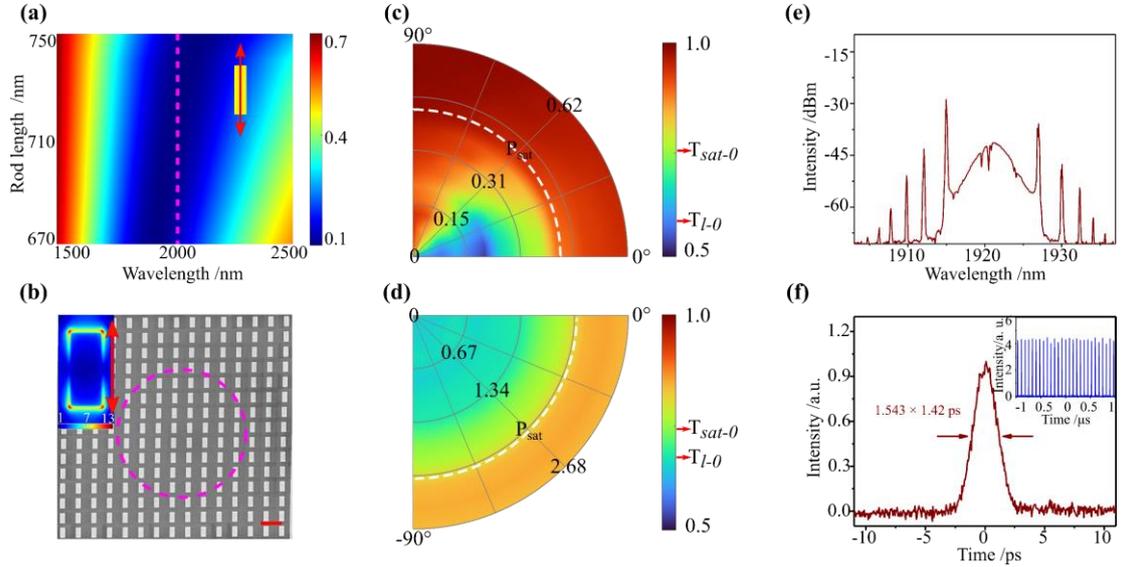

**Figure 5. Plasmonic metafiber for the soliton mode-locking at around the 2 μm optical waveband.** (a) Calculated transmission spectra of the Au nanorod metasurfaces with varying rod lengths. The rod width is 340 nm. The longitudinal and transverse periods keep constant as 1μm. The excitation polarization is parallel to the long axis of the nanorods. The height of the nanostructures is set to 60 nm. The pink dashed line indicates the spectral wavelength of 2 μm. (b) SEM image of an Au nanorod metasurface fabricated on the endface of a standard SMFJ by using FIB. The length and width of the nanorods are 690 nm and 345 nm with a tolerance of 10 nm, respectively. The longitudinal and transverse periods are the same: 1μm. The height of the nanostructures is 60 nm. The scale bar represents 1μm. The dashed pink circle indicates the boundary of the fiber core. The inset shows the fundamental electrical field distribution of the nanorod array (taking periodic conditions into account). The red arrow in the inset indicates the input polarization. (c~d) Power- and polarization-dependent nonlinear transmission of a nanorod array in on-resonance (using 1950 nm fs laser) and off-resonance (using 1550 nm fs laser) excitation conditions. The polar coordinates (P, θ) represent the averaged power in the focus and input polarization, respectively. The color contrast represents the transmission level of the nanorod array. The white dashed curves represent averaged saturation powers. $T_{l-0}$ and $T_{sat-0}$ denote the transmission levels at the initial states of linear absorption and saturable absorption when θ=0°, respectively. (e) Soliton pulse

spectrum generated by a thulium-doped fiber laser incorporating a metafiber SA, at the pump power of 310 mW. (f) Autocorrelation trace of a single soliton at the pump power of 310 mW. Inset shows the pulse trains over a longer time range (2 μs). The time interval is 72 ns.

It is notable here that we haven't managed to reach any mode-locked regime in the off-resonance cases (as shown in Fig. 2(f) and Fig. 5(d)), demonstrating the unambiguous roles of plasmonic resonances in the saturation absorption and promotion of laser mode-locking. We also observe no visible thermal damage from such metafiber SAs following to their usage in ultrafast pulse generation (for both 1.5 μm and 2 μm cases) even at the maximal pump powers, indicating the good thermal conductivity and stability of the plasmonic metasurfaces against optical damage (see "Thermal damage threshold estimation" from SI for details).

**Discussions**

We have achieved for the first time highly stable soliton mode-locking at different optical wavebands, promoted by plasmonic metasurfaces directly patterned on the fiber facets. Considering previous work on solid-state SAs, e. g. semiconductor saturable absorber mirror, [36] our metafiber SAs allow for more cost-effective, timesaving, broadband and tunable nonlinear absorption compatible with all-fiber operation for ultrashort pulse generation and reshaping. [15, 37] Also, due to the short hot carrier relaxation dynamics arising from a plasmonic system, the pulse duration from a metafiber laser can easily reach a typical sub-picosecond time scale, in contrast to the situation with common semiconductors having relaxation times of several ps. [6, 38, 39] Compared with the widely investigated low dimensional materials, including graphene,[33] graphene oxide, [40] carbon nanotubes,[41] black phosphorus [42], transition metal dichalcogenides, [43-46] topological insulators [47, 48] and colloidal nanoparticles, [14, 34, 35] in additional to fully competitive or superior performances on saturable absorption and laser mode-locking (Fig. 6(a)), the remarkable properties reported here mainly originate from the plasmonic nature of the metasurfaces themselves. Indeed, the plasmonic modes of the metasurfaces, which are highly size-, shape- and orientation-dependent, are fully guaranteed by highly resolved nanofabrication techniques coming from planar technologies. Therefore, the resonance wavelength and polarimetric properties of metasurfaces can be quantitatively tuned by design. These will allow the metasurfaces to reach an extraordinarily high modulation depth still compatible with a relatively low pump threshold for promoting mode-locked regimes. The merits of plasmonic metasurfaces compared with other 2D materials with dispersed sizes and random orientations is illustrated in Fig. 6(a). A large modulation depth leads to a reliable self-starting, due to the strong pulse shaping performed by the SAs. [49] A low mode-locking threshold is also particularly beneficial to achieve, in a second step, harmonic mode locking which multiplies the laser output repetition rate. [45] Considering the pioneering works using plasmonic metasurfaces as SAs in a free-space coupled laser cavity, [15] our metafiber will allow a straightforward implementation within standard fiber networks, targeting an ultracompact fiber laser and optical pulse reshaper, as well as other "all-in-fiber" optical systems.

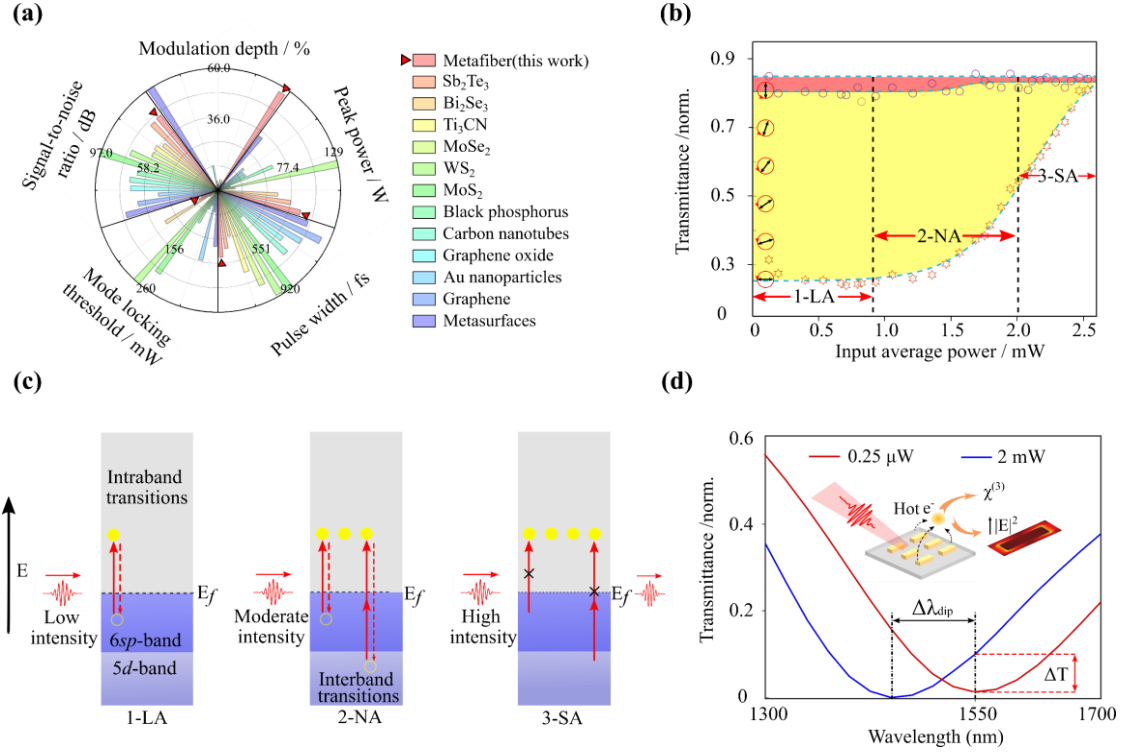

**Figure 6. Laser performance comparisons and saturable absorption mechanisms for the metafiber SAs.** (a) Performance comparisons of soliton mode-locking lasers at 1.5 μm optical waveband using different material SAs. The metafiber SA is highlighted with red triangles. (b) Power- and polarization-dependent saturation process of plasmonic metasurfaces. Three stages leading to the saturation absorptions are divided in accordance with pumping laser power: linear absorption (LA), nonlinear absorption (NA) and saturation absorption (SA). Two contributors are involved depending on the excitation polarization for each stage: bulk material effect (filled with red) and surface plasmon effect (filled with yellow). The opened stars and circles are the raw transmission data from Fig. 2(e) for the polarization angles 0° and 90° with respect to the long axis of nanorods, respectively. The dashed cyan lines are the corresponding fitted data. The black arrows indicate the excitation polarizations. (c) Saturation absorption mechanism from the aspect of the bulk material effect. Carrier transitions occur at given high density-of-states symmetry points of energy bands as the input power increases. The closed yellow circles stand for the electrons and the opened circles are the holes left. The solid and dashed arrows represent the excitation and relaxation processes, respectively. (d) Saturation absorption mechanism from the aspect of surface plasmon effect. Surface plasmon induced nonlinearity shifts the absorption resonance ($\Delta\lambda_{dip}$) and modifies the transmission level ($\Delta T$) of a plasmonic SA. The red and blue lines represent the transmission spectra of a nanorod metasurface excited with an averaged input power of 0.25 μW and 2 mW in a fs laser focus, respectively.

As material-based SAs become one of the key elements for a fiber laser architecture to achieve stable passive mode locking, it is worth clarifying the physical mechanisms related to the saturable absorption of plasmonic metasurfaces. Here below we propose several principles from both the macroscopic and microscopic points of view to better interpret the physical natures, as the classical absorption and transmission models used in SAs basically treat the bulk semiconductor materials. First, it is essential to notice the intrinsic differences between the bulk and nanoscale materials. As the size decreases to nanoscale, the properties of atoms present on the outer boundary of particles become dominant. [50] The optical properties change significantly in comparison with those of the bulk due to their size-dependent interactions with light. This works in particular for the plasmonic metasurfaces, which rely on the surface plasmon resonances contributed by collective conduction electron oscillations. [6] Following this principle, the saturable absorption of plasmonic metasurfaces could be generally divided into two contributors: the bulk material effect and the surface plasmon effect. As illustrated in Fig. 6(b), the former (filled with red color) could be extracted from a special 'off-resonance excitation' test, in which the excitation polarization is perpendicular to the long axis of nanorods. In such circumstance, the longitudinal dipolar plasmon mode of the metasurfaces, which plays the dominant role on the unusual polarimetric linear and nonlinear optics performances, will be almost entirely depressed. As can be seen from Fig. 6(b), the transmission remains a global high level (>80%) and almost immune to the increasing input power. The saturation process induced by the bulk metal materials could be still described by the classical energy band theory. The reason stems from the physical fact that there are forbidden bands at some high density-of-states symmetry points for noble metals, e. g. X or L points in the first Brillouin zone of gold, [51, 52], where most of the carrier transitions take place in a light excitation. [6, 53] As shown in Figs. 6(b, c), the whole absorption process with the increasing pumping power could be generally divided into three stages: linear absorption, nonlinear absorption and saturable absorption. We notice here that the launch of next step doesn't simultaneously quit the previous steps. In the linear absorption stage, as depicted in Fig. 6(c), the *sp* electrons below Fermi level $E_f$ are excited to the states above $E_f$ through intraband transitions. The excited electrons relax to *sp* bands and recombine with the holes left at a sub-picosecond timescale. [54, 55] In the ultrafast pumping regime, i. e. the pulse duration of the pumping laser is shorter than the relaxation time of hot electrons, the electrons from the *d* bands can also be excited to the state above $E_f$ through interband transitions, inducing multiphoton absorption, [6, 52, 53] e. g. two-photon absorption as depicted in Fig. 6(c). Such nonlinear absorption gradually dominates as the pumping intensity increases. When the pumping intensity becomes sufficient high, the photogenerated carrier population increases significantly and occupies all the states in the conduction bands, which would block further absorption due to the Pauli exclusion principle, leading to a saturated absorption of plasmonic metasurfaces. [38, 56]

Next, we make more thorough discussion on the surface plasmon effect, which contributes to the majority of saturable absorption for the plasmonic metasurfaces, as shown the yellow region in Fig. 6(b). The surface plasmons can be interpreted to play a twofold role in accordance with the light-matter interaction formular: $P(t)=\varepsilon_0\chi^{(1)}\mathbf{E}(t)+\varepsilon_0(\chi^{(2)}\mathbf{E}(t)^2+\chi^{(3)}\mathbf{E}(t)^3+…)$, where $\varepsilon_0$ is the permittivity of free space, $\chi^{(1)},\chi^{(2)}$ and $\chi^{(3)}$ are the first three terms of susceptibilities.[57] As illustrated in the inset of Fig. 6(d), on one hand, the surface plasmons provide a dramatic local field enhancement for the electric field **E**, especially when their resonances match the fundamental or high-order harmonic excitation frequencies. [10, 58, 59] On the other hand, the surface components of nonlinear susceptibilities, e. g. $\chi^{(3)}_{surf}$ in the process of two-photon absorption, will be significantly alerted due to the hot carrier redistributions during the strong light-matter interactions.[31, 38, 39, 60] Given the above two attributions, an extra hot carriers (mainly from electrons) induced nonlinearity is produced and shifts the absorption resonances [60]. As demonstrated in Fig. 6(d), considering an increase of input power from 0.25 μW to 2 mW, the plasmonic resonance initially designed at 1550 nm is shifted to 1457 nm and the transmission level shows an increase of 8.5% (see "Optical nonlinearity induced plasmonic resonance shift" from SI for details). Thus, strong dependences on the pumping laser power and polarization for the transient transmission of plasmonic metasurfaces are another key origin forming the saturated absorption.

**Conclusion**

We report on a novel nanomanufacturing method to integrate plasmonic metasurfaces directly on the fiber endfaces of commercial SMFs and a first demonstration of their applications in nonlinear optics regime. In contrast with the metasurfaces solely on bare fibers that most of current studies consider, our approach treats the SMFs or more general multi-mode fibers with universal adapting interfaces, yielding advantages of compatibility with fiber networks and robustness against external influences such as bending and mechanical vibrations. Also, as only standard nanofabrication techniques are involved, the process flow could be accessed by worldwide cleanrooms. Featuring various plug-play connections, low insertion loss and cost-effective nanofabrication, the metafibers with functionalized plasmonic properties could potentially become a new regular fiber component in the fiber optics community. To demonstrate a practical application, the metafibers are implemented into multiple fiber laser cavities, acting as special SAs to assist lasers to mode lock. Result shows that the metafiber SAs have unique features involving broad absorbing spectra, tunable plasmonic resonances, high modulation depth and ultrafast recovery time, which are competitive to the best performances of the state-of-art 2D materials. All-fiber sub-picosecond mode-locked laser pulses are generated at 1.5 μm and 2 μm regions. The application of such SAs can be immediately extended to other optical wavebands by tuning the plasmonic resonances of the metafibers. As such, saturable plasmonic metafibers provide ultrathin SAs for applications where tunable nonlinear transfer functions are needed, such as in ultrafast lasers, pulse shaping and regeneration, and become key-enablers to compact and efficient optical frequency combs or neuromorphic circuits. The work enriches the 'lab-on-fiber' paradigm and offers an avenue to realize the "all-in-fibers" optical systems in sensing, imaging and communications.

**Materials and methods**

All of the details about the nanofabrication, extinction simulations, linear and nonlinear optical characterizations, saturation absorption model, soliton mode locking at 2 μm, damage threshold test on the plasmonic metafibers and optical nonlinearity induced plasmonic resonance shift are provided in the Supplementary Information.

**Acknowledgements**

The authors gratefully acknowledge the funds from the National Natural Science Foundation of China (61927820, 51806199, 61905200) and the National Key Research and Development Program of China (2017YFA0205700). J. Wang thank Dr. Ziyang Zhang from School of Engineering, Westlake University for the technical support of fiber setups. J. Wang, L. Zhang, B. Chen and Q. Jia thank Westlake Center for Micro/Nano Fabrication for the facility support and technical assistance. This is a preprint of an article published in Light: Advanced Manufacturing. The final authenticated version is available online at: https://doi.org/10.37188/lam.2022.045

**Author contributions**

All of the authors contributed extensively to the work presented in this paper. M. Qiu and J. Wang lead the whole research project. B. Cluzel, A. Coillet, P. Grelu, J. Wang and M. Qiu conceived the main conceptual ideas. J. Wang and L. Zhang developed the methodologies to fabricate the metafibers. D. Wu and X. Shen supervised the work on the fiber laser mode-locking. L. Zhang, Q. Jia, B. Chen prepared the samples. L. Zhang, N. Tang, F. Liu, X. Sun, H. Yu, F. Qiu, X. Liu and J. Qiu contributed to the linear and nonlinear optical characterizations. H. Zhang and X. Chen conducted the fiber laser mode-locking experiments. J. Wang, L. Ying, W. Yan, W. E. I. Sha, L. Zhang and M. Qiu proposed the principles of the saturation absorption mechanisms. All of the authors contributed to the discussions and the writing of the manuscript.

**Conflict of interest**

The authors declare that they have no conflict of interest.

**Reference**

[1] Shelby, R. A.; Smith, D. R.; Schultz, S. Experimental verification of a negative index of refraction. *Science* **2001,** 292, 77-79.

[2] Schurig, D.; Mock, J. J.; Justice, B. J.; Cummer, S. A.; Pendry, J. B.; Starr, A. F.; Smith, D. R. Metamaterial electromagnetic clock at microwave frequencies. *Science* **2006,** 314, 977-980.

[3] Sun, S.; He, Q.; Xiao, S.; Xu, Q.; Li, X.; Zhou, L. Gradient-index meta-surfaces as a bridge linking propagating waves and surface waves. *Nat. Mater*. **2012**, 11, 426-431.

[4] Yu, N.; Capasso, F. Flat optics with designer metasurfaces. *Nat. Mater.* **2014**, 13, 139-150.

[5] Maier, S. A. Surface Plasmon Polaritons at Metal /Insulator Interfaces. *Plasmonics: Fundamentals and Applications*; Springer: New York, **2007**; pp 25-26.

[6] Wang, J.; Gurdal, E.; Horneber, A.; Dickreuter, S.; Kostcheev, S.; Meixner, A. J.; Fleischer, M.; Adam, P. M.; Zhang, D. Carrier recombination and plasmonic emission channels in metallic photoluminescence. *Nanoscale* **2018**, 10, 8240-8245.

[7] Krasnok, A.; Tymchenko, M.; Alù, A. Nonlinear metasurfaces: a paradigm shift in nonlinear optics. *Mater. Today* **2018**, 21, 8-21.

[8] Zhao, Y.; Yang, Y.; Sun, H. Nonlinear meta-optics towards applications. *PhotoniX* **2021**, 2, 3.

[9] Minovich, A. E.; Miroshnichenko, A. E.; Bykov, A. Y.; Murzina, T. V.; Neshev, D. N.; Kivshar, Y. S. Functional and nonlinear optical metasurfaces. *Laser Photonics Rev*. **2015**, 9, 195-213.

[10] Wang, J.; Butet, J.; Bernasconi, G. D.; Baudrion, A. L.; Leveque, G.; Horrer, A.; Horneber, A.; Martin, O.; Meixner, A. J.; Fleischer, M.; Adam, P. M.; Zhang, D. Strong second-harmonic generation from Au-Al heterodimers. *Nanoscale* **2019**, 11, 23475-23481.

[11] Shastri, B. J.; Tait, A. N.; Ferreira De Lima, T.; Pernice, W. H. P.; Bhaskaran, H.; Wright, C. D.; Prucnal, P. R. Photonics for artificial intelligence and neuromorphic computing. *Nat. Photonics* **2021**, 15, 102-114.

[12] Kumar, S.; Williams, R. S.; Wang, Z. Third-order nanocircuit elements for neuromorphic engineering. *Nature* **2020**, 585, 518-523.

[13] Goi, E.; Zhang, Q.; Chen, X.; Luan, H.; Gu, M. Perspective on photonic memristive neuromorphic computing. *PhotoniX* **2020,** 1, 3.

[14] Kang, Z.; Xu, Y.; Zhang, L.; Jia, Z.; Liu, L.; Zhao, D.; Feng, Y.; Qin, G.; Qin, W. Passively mode-locking induced by gold nanorods in erbium-doped fiber lasers. *Appl. Phys. Lett*. **2013**, 103, 041105.

[15] Wang, J.; Coillet, A.; Demichel, O.; Wang, Z.; Rego, D.; Bouhelier, A.; Grelu, P.; Cluzel, B. Saturable plasmonic metasurfaces for laser mode locking. *Light Sci. Appl*. **2020**, 9, 50.

[16] Zhao, D.; Liu, Y.; Qiu, J.; Liu, X. Plasmonic saturable absorbers. *Adv. Photonics Res.* **2021**, 2, 2100003.

[17] Wang, X.; Luo, Z.; Liu, H.; Liu, M.; Luo, A.; Xu, W. Microfiber-based gold nanorods as saturable absorber for femtosecond pulse generation in a fiber laser. *Appl. Phys. Lett.* **2014**, 105, 161107.

[18] Shu, Y.; Guo, P.; Li, X.; Li, G.; Wang, P.; Shen, G.; Li, J. Gold nanorods as saturable absorber for harmonic soliton molecules generation. *Front. Chem.* **2019**, 7, 715.

[19] Han, A.; Kuan, A.; Golovchenko, J.; Branton, D. Nanopatterning on nonplanar and fragile substrates with ice resists. *Nano Lett.* **2012,** 12, 1018-1021.

[20] Hong, Y.; Zhao, D.; Liu, D.; Ma, B.; Yao, G.; Li, Q.; Han, A.; Qiu, M. Three-dimensional in situ electron-beam lithography using water ice. *Nano Lett*. **2018,** 18, 5036-5041.

[21] Xiong, Y.; Xu, F. Multifunctional integration on optical fiber tips: challenges and opportunities. *Adv. photonics* **2020**, 2, 064001.

[22] Principe, M.; Consales, M.; Micco, A.; Crescitelli, A.; Castaldi, G.; Esposito, E.; La Ferrara, V.; Cutolo, A.; Galdi, V.; Cusano, A. Optical fiber meta-tips. *Light Sci. Appl*. **2017**, 6, e16226.

[23] Hong, Y.; Zhao, D.; Wang, J.; Lu, J.; Yao, G.; Liu, D.; Luo, H.; Li, Q.; Qiu, M. Solvent-free nanofabrication based on ice-assisted electron-beam lithography. *Nano Lett*. **2020**, 20, 8841-8846.

[24] Umakoshi, T.; Saito, Y.; Verma, P. Highly efficient plasmonic tip design for plasmon nanofocusing in near-field optical microscopy. *Nanoscale* **2016**, 8, 5634-5640.

[25] Plidschun, M.; Ren, H.; Kim, J.; Förster, R.; Maier, S. A.; Schmidt, M. A. Ultrahigh numerical aperture meta-fibre for flexible optical trapping. *Light Sci. Appl*. **2021**, 10, 57.

[26] Consales, M.; Ricciardi, A.; Crescitelli, A.; Esposito, E.; Cutolo, A.; Cusano, A. Lab-on-fiber technology: toward multifunctional optical nanoprobes. *ACS Nano* **2012**, 6, 3163-3170.

[27] Feng, S.; Darmawi, S.; Henning, T.; Klar, P. J.; Zhang, X. A miniaturized sensor consisting of concentric metallic nanorings on the end facet of an optical fiber. *Small* **2012,** 8, 1937-1944.

[28] Calafiore, G.; Koshelev, A.; Allen, F. I.; Dhuey, S.; Sassolini, S.; Wong, E.; Lum, P.; Munechika, K.; Cabrini, S. Nanoimprint of a 3D structure on an optical fiber for light wavefront manipulation. *Nanotechnology* **2016**, 27, 375301.

[29] Calafiore, G.; Koshelev, A.; Darlington, T. P.; Borys, N. J.; Melli, M.; Polyakov, A.; Cantarella, G.; Allen, F. I.; Lum, P.; Wong, E.; Sassolini, S.; Weber-Bargioni, A.; Schuck, P. J.; Cabrini, S.; Munechika, K. Campanile near-field probes fabricated by nanoimprint lithography on the facet of an optical fiber. *Sci. Rep.* **2017**, 7, 1651.

[30] Horneber, A.; Baudrion, A. L.; Adam, P. M.; Meixner, A. J.; Zhang, D. Compositional-asymmetry influenced non-linear optical processes of plasmonic nanoparticle dimers. *Phys. Chem. Chem. Phys.* **2013**, 15, 8031-8034.

[31] Wang, J.; Baudrion, A.; Béal, J.; Horneber, A.; Tang, F.; Butet, J.; Martin, O. J. F.; Meixner, A. J.; Adam, P.; Zhang, D. Hot carrier-mediated avalanche multiphoton photoluminescence from coupled Au–Al nanoantennas. *J. Chem. Phys.* **2021**, 154, 074701.

[32] Sun, Y.; Xia, Y. Shape-controlled synthesis of gold and silver nanoparticles. *Science* **2002**, 298, 2176-2179.

[33] Fu, B.; Hua, Y.; Xiao, X.; Member.; IEEE.; Zhu, H.; Sun, Z.; Yang, C. Broadband graphene saturable absorber for pulsed fiber lasers at 1, 1.5, and 2 μm. *IEEE J. Sel. Top. Quantum Electron*. **2014**, 20, 5.

[34] Agrawal, G. P. Optical Fibers, Control of Nonlinear Effects. *Fiber-Optics Communication Systems*; John Wiley & Sons: New York, **2010**; pp 47-48, 416-417.

[35] Kelly, S. M. J. Characteristic sideband instability of periodically amplified average soliton. *Electron. Lett.* **1992**, 28, 806-808.

[36] Keller, U. Recent developments in compact ultrafast lasers. *Nature* **2003**, 424, 831-838.

[37] Liu, W.; Pang, L.; Han, H.; Bi, K.; Lei, M.; Wei, Z. Tungsten disulphide for ultrashort pulse

generation in all-fiber lasers. *Nanoscale* **2017**, 9, 5806-5811.

[38] Fatti, D. N.; Voisin, C.; Achermann, M.; Tzortzakis, S.; Christofilos, D.; Vallee, F. Nonequilibrium electron dynamics in noble metals. *Phys. Rev. B.* **2000**, 61, 16956-16966.

[39] Baida, H.; Mongin, D.; Christofilos, D.; Bachelier, G.; Crut, A.; Maioli, P.; Del, F. N.; Vallee, F. Ultrafast nonlinear optical response of a single gold nanorod near its surface plasmon resonance. *Phys. Rev. Lett*. **2011**, 107, 057402.

[40] Boguslawski, J.; Sotor, J.; Sobon, G.; Kozinski, R.; Librant, K.; Aksienionek, M.; Lipinska, L.; Abramski, K. M. Graphene oxide paper as a saturable absorber for Er- and Tm-doped fiber lasers. *Photonics Res*. **2015**, 3, 119.

[41] Wang, F.; Rozhin, A. G.; Scardaci, V.; Sun, Z.; Hennrich, F.; White, I. H.; Milne, W. I.; Ferrari, A. C. Wideband-tuneable, nanotube mode-locked, fibre laser. *Nat. Nanotechnol*. **2008**, 3, 738-742.

[42] Sotor, J.; Sobon, G.; Macherzynski, W.; Paletko, P.; AbramskiBP, K. M. Black phosphorus saturable absorber for ultrashort pulse generation. *Appl. Phys. Lett*. **2015**, 107, 051108.

[43] Wu, K.; Zhang, X.; Wang, J.; Chen, J. 463-MHz fundamental mode-locked fiber laser based on few-layer $MoS_2$ saturable absorber. *Opt. Lett*. **2015**, 40, 428-431.

[44] Wu, K., Zhang, X., Wang, J., Li, X., & Chen, J. $WS_2$ as a saturable absorber for ultrafast photonic applications of mode-locked and Q-switched lasers. *Opt. Express* **2015**, 23, 9.

[45] Koo, J.; Park, J.; Lee, J.; Jhon, Y. M.; Lee, J. H. Femtosecond harmonic mode-locking of a fiber laser at 3.27 GHz using a bulk-like, $MoSe_2$-based saturable absorber. *Opt. Express* **2016**, 24, 10.

[46] Jhon, Y. I.; Koo, J.; Anasori, B.; Seo, M.; Lee, J. H.; Gogotsi, Y.; Jhon, Y. M. Metallic MXene saturable absorber for femtosecond mode-locked lasers. *Adv. Mater*. **2017**, 29, 1702496.

[47] Li, K.; Song, Y.; Yu, Z.; Xu, R.; Dou, Z.; Tian, J. L-band femtosecond fibre laser based on $Bi_2Se_3$ topological insulator. *Laser Phys. Lett*. **2015**, 12, 105103.

[48] Boguslawski, J.; Sotor, J.; Sobon, G.; Tarka, J.; Jagiello, J.; Macherzynski, W.; Lipinska, L.; Abramski, K. M. Mode-locked Er-doped fiber laser based on liquid phase exfoliated $Sb_2Te_3$ topological insulator. *Laser Phys*. **2014**, 24, 105111.

[49] Cheng, L.; Yuan, Y.; Liu, C.; Cao, X.; Su, J.; Zhang, X.; Zhang, H.; Zhao, H.; Xu, M.; Li, J. Linear and nonlinear optical properties modulation of $Sb_2Te_3$/GeTe bilayer film as a promising saturable absorber. *Results Phys*. **2019**, 13, 102282.

[50] Pareek, V.; Bhargava, A.; Gupta, R.; Jain, N.; Panwar, J. Synthesis and applications of noble metal nanoparticles: a review. *Adv. Sci. Eng. Med.* **2017**, 9, 527-544.

[51] Guerrisi, M.; Rosei, R.; Winsemius, P. Splitting of the interband absorption edge in Au. *Phys. Rev. B* **1975**, 12, 557-563.

[52] Boyd, G. T.; Yu, Z. H.; Shen, Y. R. Photoinduced luminescence from the noble metals and its enhancement on roughened surfaces. *Phys. Rev. B* **1986**, 33, 7923-7936.

[53] Imura, K.; Nagahara, T.; Okamoto, H. Near-field two-photon-induced photoluminescence from single gold nanorods and imaging of plasmon modes. *J. Phys. Chem. B* **2005**, 109, 13214-13220.

[54] Wang, J.; Butet, J.; Baudrion, A.; Horrer, A.; Lévêque, G.; Martin, O. J. F.; Meixner, A. J.; Fleischer, M.; Adam, P.; Horneber, A.; Zhang, D. Direct comparison of second harmonic generation and two-photon photoluminescence from single connected gold nanodimers. *J. Phys. Chem. C* **2016**, 120, 17699-17710.

[55] Hohlfeld, J.; Wellershoff, S. S.; Güdde, J.; Conrad, U.; Jähnke, V.; Matthias, E. Electron and lattice dynamics following optical excitation of metals. *Chem. Phys*. **2000**, 251, 237-258.

[56] Schroeder, W. U.; Birkelund, J. R.; Huizenga, J. R; Wilcke, W. W.; Randrup, J. Effect of Pauli

blocking on exchange and dissipation mechanisms operating in heavy-ion reactions. *Phys. Rev. Lett.* **1980**, 44, 308-312.

[57] Boyd, R. W. The Nonlinear Optical Susceptibility. *Nonlinear Optics*; Academic press: New York, **2008**; pp 1-33

[58] Thyagarajan, K.; Butet, J.; Martin, O. J. F. Augmenting second harmonic generation using Fano resonances in plasmonic systems. *Nano Lett.* **2013**, 13, 1847-1851.

[59] Ren, M.; Liu, S.; Wang, B.; Chen, B.; Li, J.; Li, Z. Giant enhancement of second harmonic generation by engineering double plasmonic resonances at nanoscale. *Opt. Express* **2014**, 22, 28653.

[60] Hache, F.; Ricard, D.; Flytzanis, C.; Kreibig, U. The optical kerr effect in small metal particles and metal colloids: The case of gold. *Appl. Phys. A* **1988**, 47, 347-357.

Supporting information

**Plug-Play Plasmonic Metafibers for Ultrafast Fiber Lasers**

*Corresponding authors: qiumin@westlake.edu.cn; wangjiyong@westlake.edu.cn; wuduanduan@nbu.edu.cn; shenxiang@nbu.edu.cn

# Content

# 1. Nanofabrication of the metafiber by using EBL

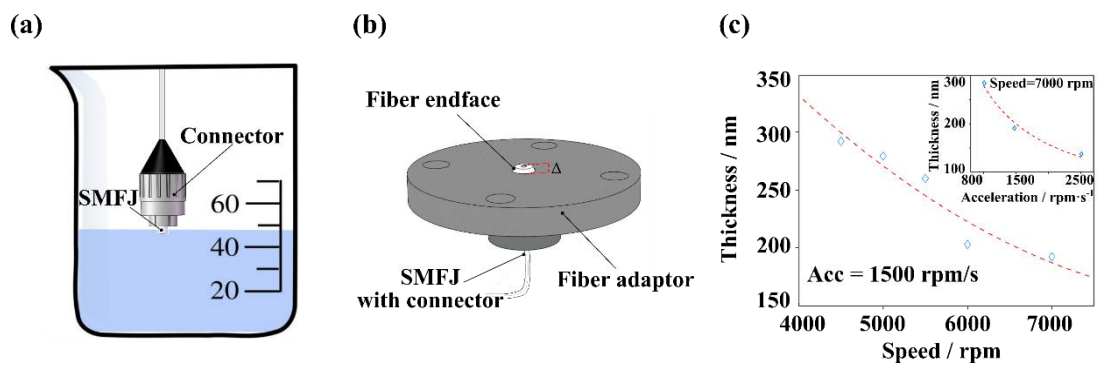

Figure S1. The schematic process flow of nanofabrication on the fiber tips of SMJPs using EBL. (a) Pre-cleaning a SMFJ tip. (b) Home-built fiber adapter. (c) The thickness of electron-sensitive resist as functions of spin-coating speed and the acceleration (inset). The blue diamonds and the dashed red curves are the experimental and fitted data, respectively.

To fabricate the well-patterned metasurfaces on the endface of a commercial single-mode fiber jumper (SMFJ) by using electron-beam lithography (EBL), standard procedures including cleaning, spin-coating, electron-beam exposure, development, physical vapor deposition and lift-off are performed. A SMFJ is cut into two identical parts, each of which maintains a standard connector in one end. The fiber tips with the connector of SMFJs are dipped into chemical solvents of a beaker (Fig. S1(a)), and then sequentially cleaned by acetone, isopropanol and deionized water in an ultrasonic sink for 10 min for each step. The fiber tips are dried by nitrogen flows and then mounted to a coaxial fiber adaptor (Fig. S1(b)), while the other end without the connector passes through a hole drilled at the bottom of the rotating /translating chunk and twines around the body, as illustrated in Fig. 1(b1-b3). The top surface of the fiber adaptor is designed slightly lower than the SMFJ endface, so that the spreading resist isn't bounced back once it touches the tip-adaptor boundary during the spin-coating. The fiber adaptor can be connected with a home-made rotating chunk via four screws, as shown in Fig. 1(b1) in the main text. The bottom part of the rotating chunk is designed to adapt a commercial spin-coater (SUSS MicroTec). 30 uL of electron-sensitive resist (PMMA 950K, Allresist) is dropped on the fiber tip. In accordance with our empirical database (Fig. S1(c)), the following parameters for spin-coating are prudently used to obtain a 200 ± 5 nm thickness of electron resist: speed-7000 rpm and acceleration- 1500 rpm/s. The rotating chunk is then flipped top down and placed on a hot plate for soft baking (175 °C for 3 min), as shown in Fig. 1(b2) in the main text. The spacing between the fiber endface and the top surface of the hot plate is guaranteed by the screws' heads: 2.5 mm. The fiber tip is then spin-coated with a thin layer of conductive polymer (~ 40 nm, AR-PC 5090.02, Allresist) to weaken the charge accumulations during the electron-beam observations or exposures. The fiber adaptor is removed from the rotating chunk to a home-built translating chunk, which is adapted to the scanning stage of a commercial SEM (Zeiss Crossbeam 550), as shown in Fig. 1(b3) in the main text. After the patterning, the fiber is unloaded from the fiber adaptor and the conductive polymer is first removed by dipping the fiber tips into deionized water. The optical fiber is then sequentially dipped into the developer (Allresist, 1 min) and stopper (Allresisit, 1min) for the development. The well-patterned SMFJ is then transferred to a home-built evaporation adaptor for depositing the target materials (3 nm Cr and 50 nm Au), as illustrated in Fig. 1(b4) in the main text. After remaining the SMFJ into the acetone for more than 24 h, we finally obtain the plasmonic metasurfaces (Au nanorods array) on the SMFJ end facet following a standard lift-off process.

## 2. Nanofabrication of the metafiber by using FIB

The fiber tips of SMFJs are cleaned and dried in accordance with the procedures described above. They are then mounted on a home-built evaporation adaptor, as illustrated in Fig. 1(b4) in the main text. The evaporation adaptor is then adapted to the sample holder of a commercial evaporator (ULVAC ei-5z) for depositing the target materials (3 nm Cr and 60 nm Au). The coated SMFJs are then removed from the evaporation adaptor to the home-built fiber adaptors. A home-built translating chunk connecting the fiber adaptor is adapted to the electrically driving stage of a commercial FIB apparatus (Carl Zeiss, ORION NanoFab, USA) for the milling, as illustrated in Fig. 1(c1) in the main text. There are two kinds of ionized beams in the FIB chamber: $He^+$ and $Ga^+$, in which $He^+$ beam is used for the imaging and the $Ga^+$ beam is for the milling. There is an angle of 54° between the two beam columns. During the milling, the translating chunk is tilted an angle of 54° in horizontal direction for a normal milling of the $Ga^+$ beam, as illustrated in Fig. 1(c1) in the main text.

## 3. Linear optical response calculations

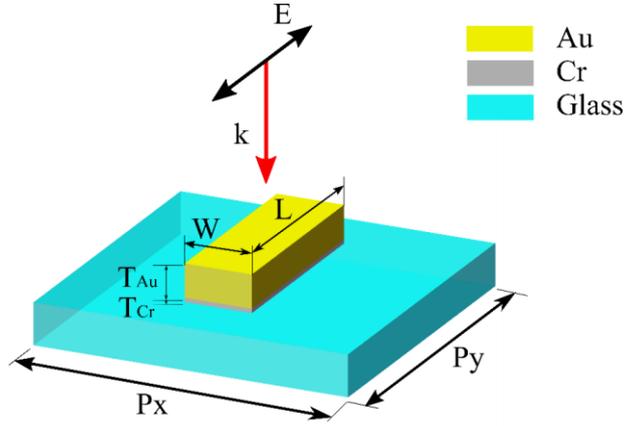

**Figure S2. Optical model of a metasurface unit cell excited with a linearly polarized light.** $P_x$ and $P_y$ denote the transverse and longitudinal periods of the nanorod fabricated on the fiber tip. $T_{Au}$ and $T_{Cr}$ represent the thickness of gold and chromium, respectively. $W$ and $L$ denote respectively the width and length of the nanorod. $k$ and $E$ represent the propagating direction of the plane wave and the polarization, respectively.

In order to calculate the linear optical responses (transmission, reflection, absorption, extinction and the fundamental electromagnetic field distributions), we use the finite-difference time-domain (FDTD) method to model the plasmonic metasurfaces. In such model, as shown in Fig. S2, the Au nanorod and Cr adhesion layer are located on a glass substrate. The refraction index of Au, Cr and glass are obtained from the material library.[1] The nanorod is excited by a linearly polarized plane wave coming from the air side ($n_{air}$=1) of the interface. The $k$ vector of the incident light is normal to the substrate plane and the polarization is parallel to the long axis of nanorod. The periodic conditions in both x and y directions are taken into account (Fig. 2(a) and Fig. 5(a) in the main text). The fundamental electric fields in each excitation wavelength (insets of Fig. 2(b) and Fig. 5(b) in the main text) are obtained simultaneously in the calculation process for each dimension of nanorod array.

## 4. Optical setups for measuring the extinction spectra

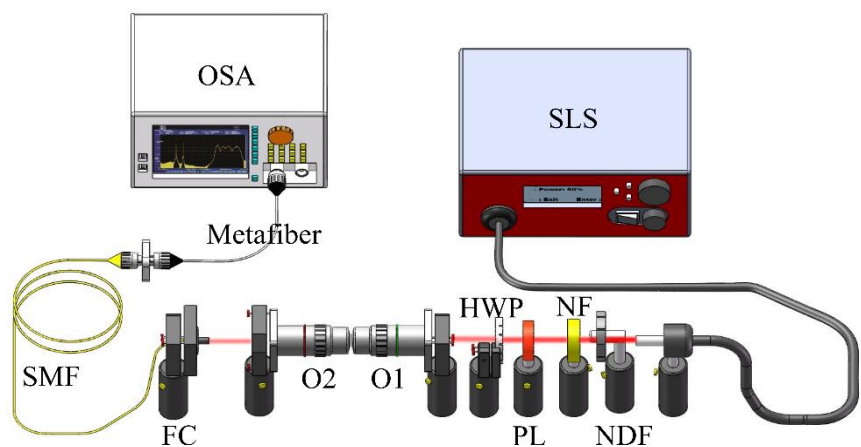

**Figure S3. The optical setups for measuring the polarimetric extinction spectra of a metafiber.** SLS is supercontinuum light source, NDF is neutral density filter, NF is notch filter, PL is polarizer, HWP is half wavelength plate, O1, 2 are objective lens, FC is fiber-optic collimator, SMF is single-mode fiber and OSA is optical spectrum analyzer.

A supercontinuum light source (SLS, NKT photonics) emitting in the 1–1.8 μm spectral range is firstly filtered via a neutral density filter (NDF) and a notch filter (NF, central wavelength: 1064 nm, band width: 44 nm) to prevent the metafiber from potential thermal damages induced by the pumping laser (1064 nm). A polarizer (PL) and a half-waveplate (HWP) control the polarization axis of the incident light, and two confocal objective lens (O1:2.5/0.08; O2: 20/0.4) changes the beam diameter to fit the fiber-optic collimator (FC) entrance. Before coupling the incident light to the testing fibers, it is necessary to introduce an intermediate single-mode fiber (SMF) to increase the mobility of the optical spectrum analyzer (OSA) and minimize the potential misalignments of the optical path in the free space during exchanging the testing fibers. The transmitted optical signals are sent to the OSA for spectrum analysis.

## 5. Optical setups for measuring nonlinear optical transmissions

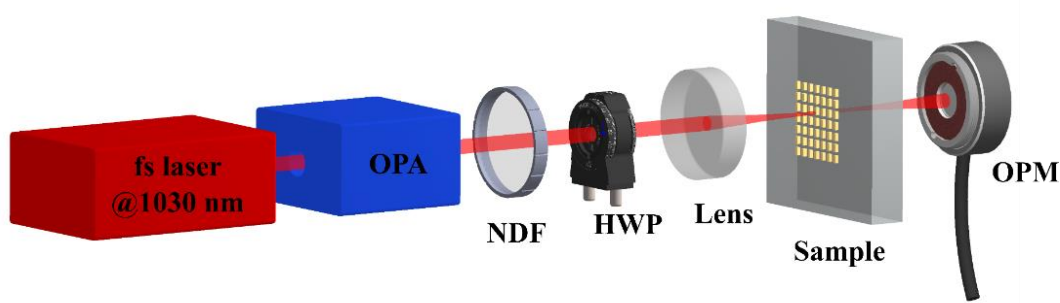

**Figure S4. Optical setups for measuring the nonlinear transmission of a nanorod metasurface.** OPA is the optical parametric amplifier, NDF is neutral density filter, HWP is half-wavelength plate and OPM is optical power meter.

Femtosecond laser pulses (pulse duration 130 fs) with a central wavelength of 1030 nm is firstly converted to other wavelengths via an optical parametric amplifier (OPA, Twin Starzz, Fastlite). A neutral density filter (NDF) is used to tune the single pulse energy and thus control the averaged power of incident light. The polarization is linear and can be further tuned by using a half-wavelength plate (HWP, 1100 nm-2000 nm, FBR-AH3c, Thorlabs). The polarization angle 0° on the sample is firstly judged by aligning the polarization with a marker which is parallel to the nanorod long axis, and is fine-tuned by observing the minimum transmission levels after rotating the HWP 360°. A lens (75 mm focal length) is used to focus the beam on the patterns of metasurfaces. The transmitted light is collected by an optical power meter (OPM, 700 nm - 1800 nm, S122C; 1200 nm - 2500 nm, S148C, Thorlabs).

## 6. Laser intensity dependent optical transmission model

The optical transmission of plasmonic metasurfaces can be generally modeled by a two-port symmetry system.[2] As sketched in Fig. S5, given an incident light $I_0$, the transmittance, reflectance and the absorbance are defined as $I_t$, $I_r$ and $I_{abs}$ respectively. When the metasurface is excited in resonant cases, e. g. fundamental wavelength matches one of plasmonic modes, the metasurface can be regarded as a resonator, which radiates in forward ($I_t$) and backward ($I_r$) directions almost equally. Assume $I_t=I_r$, then the transmission $T= I_t/I_0=(I_0-I_{abs}-I_r)/I_0$. Thus, we have $T=(I_0-I_{abs})/2I_0$. Given an effective absorption $A_{eff}$, the transmission is gained:

$$T = \frac{(1 - A_{eff})}{2} \quad (S1)$$

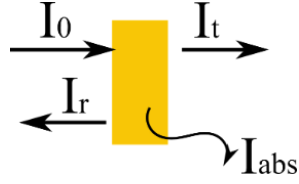

**Figure S5. Two-port symmetry model for optical transmission of plasmonic metasurfaces.** $I_0$, $I_t$, $I_r$ and $I_{abs}$ indicate the incident light, the transmittance, the reflectance and the absorbance, respectively.

The effective absorption is given by $A_{eff} = \alpha_{eff}L$, where $L$ is the penetrating distance of the incident light and $\alpha_{eff}$ is the effective absorption coefficient, which can be regarded as a combination of the linear absorption and nonlinear absorption evens, e. g. two- and three-photon absorption. Thus, the effective absorption coefficient is given by

$$\alpha_{eff} = \alpha_0 + \beta I_0 + \gamma I_0^2 + \cdots \quad (S2)$$

where $\alpha_0$, $\beta$ and $\gamma$ are the linear, two-photon and three-photon absorption coefficients, respectively. $\beta$ is also commonly called nonlinear absorption coefficient for simplification, which is connected with the both imaginary part and real part of the third-order susceptibility.[3]

Now we consider the saturation absorption in a relatively intense light, which can be modeled as

$$\alpha'_{eff} = \frac{\alpha_s}{1+I_0/I_{sat}} + \alpha_{ns} \quad (S3)$$

where $\alpha_s$ and $\alpha_{ns}$ represent saturation loss and unsaturated loss, respectively.[4] $I_{sat}$ is the saturated intensity. Considering the saturation absorption is generally induced by both linear and nonlinear optical effects, then (S3) can be further written as:

$$\alpha'_{eff}(I_0) = \frac{\alpha_0 + \beta I_0 + \gamma I_0^2 + O(I_0^3)}{1+I_0/I_{sat}} + \alpha_{ns} \quad (S4)$$

Universal intensity-dependent optical absorptions (transmissions) could be basically described by a generalized Fermi-Dirac (sigmoid) function combining both the unsaturation and saturation absorption processes:

$$T(x) = a + \frac{b}{1 + c \cdot e^{-dx}} \quad (S5)$$

The motivation mainly origins from the fact that a generalized Fermi-Dirac function is mathematically equivalent to expression (S4) but offers straightforward characteristic parameters for saturation absorption, including linear absorption coefficient, saturated intensity and modulation depth. Indeed, as shown in Fig. S6, the experimental transmission spectrum (opened stars) is fitted with the function (S5) (dashed cyan curve). The linear transmission level $T_l$ and the modulation depth $M_d$ can be extracted directly from the parameters of $a$ and $b$, respectively. The saturation power $P_{sat}$ corresponds to the transmission level $T_{sat} = T_l + M_d/2$ and unsaturated transmission is obtained by ($1-T_l-M_d$).

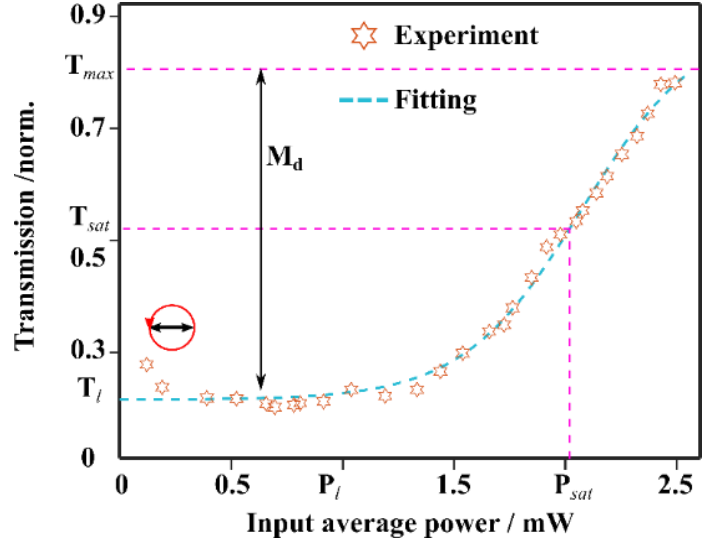

**Figure S6. Nonlinear transmission spectrum and the numerical fitting**

Here below we explain the reason why a generalized Fermi-Dirac function is mathematically equivalent to the expression (S4).

First, (S4) can be expressed as:

$$\alpha'_{eff}(x) = \frac{1}{1+x}\left(\alpha'_0 + \beta'x + \gamma'x^2 + O(x^3)\right) + \alpha_{ns} \text{ for } x=I_0/I_{sat},$$

and can be further generalized as:

$$\alpha'_{eff}(x) = \alpha''_0 + \beta''x + \gamma''x^2 + O(x^3) \tag{S6}$$

with the known Fourier expansion near zero $1/(1+x) = 1 - x + x^2 - x^3 + O(x^4)$.

Similarly, given the Fourier expansion of a normalized Fermi-Dirac function $\sigma(x)=1/(1+e^{-x})$ $=1/2+x/4 -x^3/48 + O(x^4)$, the generalized form $T(x)= a+ b/(1+c*e^{-dx})$ can thus be expanded as $T(x) = \alpha'_0 + \beta'x + \gamma'x^2 + O(x^3)$, with only a difference of a scaling factor for each term from (S6).

In most cases, the saturation absorption coefficient in (S3) is simplified as

$$\alpha'_{eff} = \frac{\alpha_0}{1 + I_0/I_{sat}} + \alpha_{ns} \tag{S7}$$

since the linear absorption normally dominates the saturation loss ($\alpha_s \approx \alpha_0$). We notice here that the generalized Fermi-Dirac function $T(x)$ is still valid to describe such saturation absorption behaviors. The normalized Fermi-Dirac function $\sigma(x)$ can be simplified by only remaining the first two terms of Fourier series of the denominator

$$\sigma(x) = \frac{1}{1+e^{-x}} = \frac{1}{1+\sum_{k=0}^{\infty}\frac{(-x)^k}{k!}} = \frac{1}{1+1-x+O(x^2)} \approx \frac{1}{2-x} \text{ for } x \in [-1,1]$$

Thus, $T(x)$ can be expressed in a similar form as equation (S7) as

$$T(I_0/I_{sat}) = \frac{\alpha'_0}{1 + I_0/I_{sat}} + \alpha'_{ns}$$

## 7. Soliton mode-locking at 2 μm

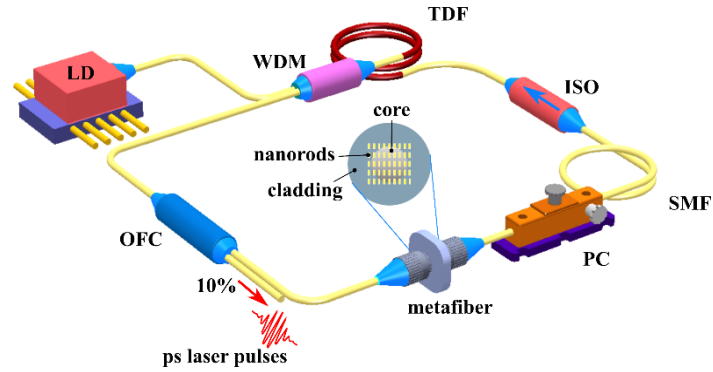

**Figure S7. Scheme of a home-built ultrafast thulium-doped fiber laser integrating a metafiber SA**, where LD represents the laser diode, WDM is the wavelength-division multiplexer, TDF is the thulium-doped fiber, ISO is the optical isolator, OFC is the output fiber coupler, SMF is the single-mode fiber, PC is the polarization controller.

A fiber cavity shown in Fig. S7 is built, which includes a 1550 nm pumping laser diode (LD), a 1560/2000 nm wavelength multiplexer (WDM), 10 cm thulium doped fiber (TDF, SM-TSF-5/125), a polarization-insensitive isolator (ISO), a polarization controller (PC), and an output fiber coupler (OFC). The coupler extracts 10% of the laser energy for the pulse characterizations. The overall length of the cavity is 14.9 m with anomalous net chromatic dispersion, and all of the fiber connections are made with standard single-mode fibers (SMFs, 11.5m long). An InGaAs PIN detector (Newport, 818-BB-51F) is employed to monitor the output signal.

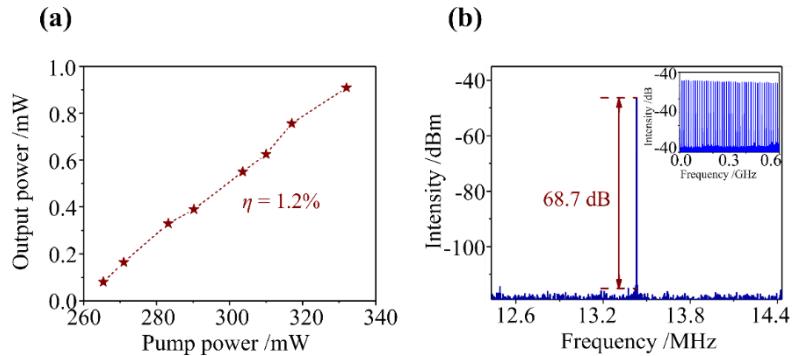

**Figure S8. Soliton mode-locking at 2 μm of a home-built fiber laser integrating a plasmonic metafiber.** (a) Averaged output power of a home-buit 2 μm fiber laser as functions of the pump power. (b) RF spectra of the soliton pulses with a scanning range of 2 MHz and a resolution of 300 Hz. Inset shows the RF spectra of the soliton pulses with a scanning range of 0-0.6 GHz and a resolution of 10 KHz.

When the metafiber is implemented in the fiber laser cavity, stable mode locking is achieved by increasing the pump power and tuning the polarization controller. Fig. S8(a) shows the laser conversion efficiency is ~1.2% for the metafiber. The pump power of a stable mode-locked pulse trains starts from 265 mW. Fig. S8(b) shows the RF spectra of the soliton mode-locked pulse at the pump power of 310 mW. The signal-to-noise ratio of the fundamental frequency of the laser reaches 68.7 dB at a resolution of 300 Hz, and it is greater than 45 dB in the range of 0-0.6 GHz at a resolution of 300 kHz, indicating a high stability of the mode locking.

## 8. Thermal damage threshold estimation

When the mode-locked pulses (pulse duration 513 fs, repetition rate 22.7 MHz) were generated, the maximum averaged power in our fiber laser cavity was 1.16 mW (when the pump power reaches to 88 mW). The maximum laser fluence inside the laser cavity can be calculated as 0.08 mJ/cm$^{-2}$. From the literature [5-7], the damage threshold of gold nanorods is within a 2.5-5.6 mJ/cm$^{-2}$ range, which is far beyond the maximum value of laser fluence in our laser cavity. Therefore, no visible thermal damages were observed from SEM images after all the mode-locking experiments.

## 9. Optical nonlinearity induced plasmonic resonance shift

Power dependent transmission spectra for the plasmonic metasurface are calculated by using finite element method (COMSOL Multiphysics). The model system is shown in Fig. S9(a), taking the periodic conditions into account.

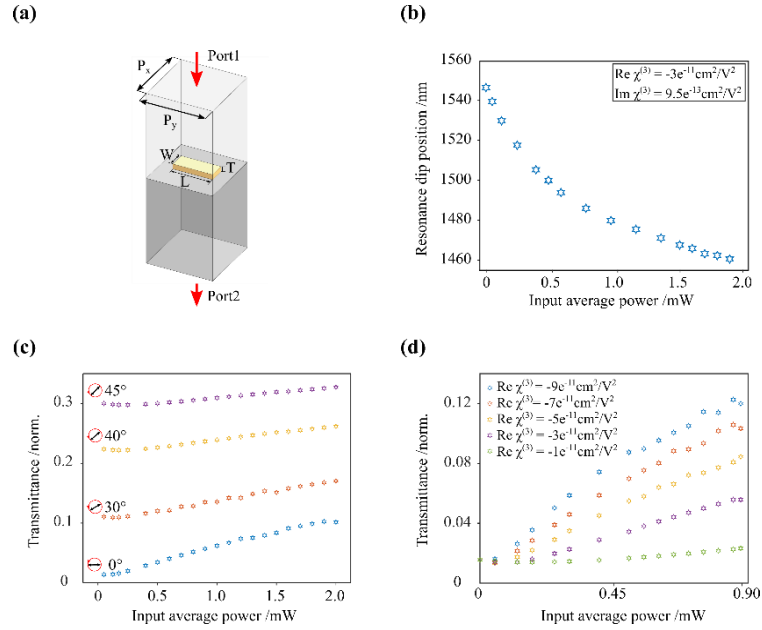

**Figure S9. Theoretical calculations on the power-dependent transmission spectra of a nanorod metasurface.** (a) 3D optical model system for the calculation of power-dependent transmission. $P_x$ and $P_y$ are the transverse and longitudinal periods of an Au nanorod array, respectively. L, W and T denote respectively the length, the width, and the thickness of the nanorods. (b) Optical nonlinearity induced dip position shifts of transmission spectra of the nanorod array. (c) Power- and polarization-dependent transmission of the nanorod array in on-resonance (optical wavelength 1550 nm) excitation condition. The black arrows represent the polarization angles with respect to the long axis of nanorods. (d) Power-dependent transmission of the nanorod array in the low input power regime as functions of Re$\chi^3$.

In such a unit cell, an Au nanorod with the length L=470 nm, width W=160 nm and the thickness T=60 nm locates on a substrate of refractive index n = 1.5, excited by a linearly polarized plane wave coming from the air side of the interface. The longitudinal and transverse periods keep the same: $P_x=P_y$=750 nm. Port 1,2 are input and output light ports, respectively. The peak power $P_{in}$ for the input port is calculated from experimental laser parameters in accordance with the following formula:

$$P_{in} = \frac{P_k}{N} = \frac{P_{avg}/(f \cdot \tau)}{(d/P_x)^2} \tag{S8}$$

where $P_k$ is the laser peak power, $N$ is the number of unit cell in the beam focus, $P_{avg}$ is the averaged power in the focus, $f$ is the repetition rate, $\tau$ is the pulse duration, $d$ is the beam diameter. The laser parameters in our experiment are listed in Table S1.

Table S1. Laser parameters used in the experiment

| Laser parameter | value |
|---|---|
| $\lambda$ | 1550 nm |
| $P_{avg}$ | 0-2.5 mW |
| $f$ | 10.32 KHz |
| $\tau$ | 179 fs |
| $d$ | ~85 μm |

In order to quantify the plasmonic resonance shift promoted by the high laser power, the optical nonlinearity dominated by the Kerr effect is considered. The relative permittivity of Au nanorods is expressed as:

$$\varepsilon = \varepsilon_L + \frac{3}{2}(Re\chi^3 - iIm\chi^3)E_{int}^2 \quad (S9)$$

where $\varepsilon_L$ is the linear relative permittivity, $E^2_{int}$ is the square of internal electric filed, $\chi^3$ is the third-order susceptibilities, $Re\chi^3$ and $Im\chi^3$ are the real part and imaginary part of $\chi^3$. Dispersed $\varepsilon_L$ is obtained from literature [1]. We notice that the $\chi^3$ values from Au films should not be used, as there are significant differences between the bulk materials and the nanostructures. Referred from the literature [3], the absolute value of $Re\chi^3$ of plasmonic nanostructures is in the orders of $10^{-10}$-$10^{-12}$ cm$^2$V$^{-2}$ and the absolute value of $Im\chi^3$ is in the orders of $10^{-10}$-$10^{-13}$ cm$^2$V$^{-2}$, which strongly depend on the excitation wavelength and the polarization. For instance, the absolute values of $Re\chi^3$ and $Im\chi^3$ are almost two orders smaller for the off-resonance situation compared with the plasmon-resonant excitation wavelength.[3] In the following calculations, initial values of $Re\chi^3$ and $Im\chi^3$ are respectively set similarly to [3]: -3×10$^{-11}$ cm$^2$V$^{-2}$ and 9.5×10$^{-13}$ cm$^2$V$^{-2}$ and later we will test how significant the $\chi^3$ values influent our results. The transmission spectra in each input power are obtained by sweeping the optical wavelength from 1300 nm to 1700 nm, as it is shown in Fig. 6(d) in the main text. The transmission dip positions are extracted from the spectra and plotted in Fig. S9(b) as a function of averaged input power. The dip position shows an obvious blue shift as the input power increases, e. g. the dip position is shifted to 1457 nm from initial 1550 nm when the input power increases from 0.25 μW to 2 mW.

Focusing on the optical wavelength of 1550 nm, we calculate the power- and polarization-dependent transmission for the nanorod array as well. As shown in Fig. S9(c), for each polarization angle $\theta$, the transmission level shows a 'S'-shape profile, which agrees with the experimental results shown in Fig. 6(b) in the main text. When $\theta$ changes from 0° to 45° with respect to the long axis of the nanorods, the transmission level in the low power regime (1-LA stage) increases due to a reduce of linear absorption of nanorod array. The modulation depth decreases as the $\theta$ increases from 0° to 45°, which generally agrees with the results shown in Fig. 2(d) in the main text.

However, we also observe that the maximum modulation depth in this model is less than 10%, which is much smaller than the experimental results shown in Fig. 2(d) and Fig. 6(b) in the main text. One of potential reasons could be that the $\chi^3$ values of plasmonic metasurfaces are power dependent, rather than constants. Fig. S9(d) shows an example how the $\chi^3$ values influence the transmission level of the nanorod array in the lower input power regime (0-0.9 mW). As the absolute value of $Re\chi^3$ increases, the transmission at the same input power increases. Following the similar tendency, the transmission in the nonlinear absorption (2-NA) and saturated absorption (3-SA) stages should reach higher levels as well, resulting in a larger modulation depth in the end. Thus, to better interpret the experimental results, the absolute values of $Re\chi^3$ should first increase with the input laser power and finally approach a constant. Considering the tight connections between the absorption coefficients and $\chi^3$ of plasmonic nanoparticles, [8] the $Re\chi^3$ should follow a similar 'S'-shaped profile defined in (S4) or (S5). These assumptions need further in-depth theoretical and experimental demonstrations, taking the contributions from the intraband transitions, interband transitions and the hot electrons into account,[8] which is, however, far beyond the scope of this work.

**Reference**

[1] Philip, H. R. Silicon Dioxide (SiO$_2$) (Glass). *Handbook of Optical Constants of Solids*; Palik E. D., Eds.; Academic press: New York, **1998**; pp 749.

[2] Fan S.; Suh W.; Joannopoulos J. D. Temporal coupled-mode theory for the Fano resonance in optical resonators. *J. Opt. Soc. Am. A* **2003**, 20, 569-572.

[3] Menezes, L. D. S.; Acioli, L. H.; Maldonado, M.; Naciri, J.; Charipar, N.; Fontana, J.; Rativa, D.; de Araújo, C. B.; Gomes, A. S. L. Large third-order nonlinear susceptibility from a gold metasurface far off the plasmonic resonance. *J. Opt. Soc. Am. B* **2019**, 36, 1485.

[4] Zhao, D.; Liu, Y.; Qiu, J.; Liu, X. Plasmonic saturable absorbers. *Adv. Photonics Res.* **2021**, 2, 2100003.

[5] Huang W.; Qian W.; El-Sayed M. A. Gold nanoparticles propulsion from surface fueled by absorption of femtosecond laser pulse at their surface plasmon resonance. *J. Am. Chem. Soc.* **2006**, 128, 13330-13331.

[6] Taylor A. B.; Siddiquee A. M.; Chon J. W. M. Below melting point photothermal reshaping of single gold nanorods driven by surface diffusion. *ACS Nano* **2014**, 8, 12071-12079.

[7] Zijlstra P.; Chon J. W. M.; Gu M. White light scattering spectroscopy and electron microscopy of laser induced melting in single gold nanorods. *Phys. Chem. Chem. Phys.* **2009**, 11, 5915.

[8] Hache F.; Ricard D.; Flytzanis C.; Kreibig U. The optical kerr effect in small metal particles and metal colloids: The case of gold. *Appl. Phys. A* **1988**, 47, 347-357.